\documentclass[epj,12pt]{svjour}
\usepackage{graphicx}
\begin{document}
\title{Transverse momentum distribution with radial flow in relativistic diffusion model}
\author{N. Suzuki\inst{1}\thanks{\emph{e-mail:}suzuki@matsu.ac.jp}
  \and M. Biyajima\inst{2}\thanks{\emph{e-mail:}biyajima@azusa.shinshu-u.ac.jp} 
}                     
%
%
\institute{Department of Comprehensive Management, Matsumoto University, Matsumoto 390-1295, Japan \and Department of Physics, Shinshu University, Matsumoto, 390-8621, Japan}
\date{Received: date / Revised version: date}
%
\abstract{
Large transverse momentum distributions of identified particles observed at RHIC are analyzed  
by a  relativistic stochastic model in the three dimensional (non-Euclidean) rapidity space. 
A distribution function obtained from the model is Gaussian-like in radial rapidity.  
It can well describe observed transverse momentum $p_T$ distributions.  
Estimation of radial flow is made from the analysis of $p_T$ distributions for $\bar{p}$ in Au + Au Collisions. 
Temperatures are estimated from observed large $p_T$ distributions under the assumption 
that the distribution function approaches to the Maxwell-Boltzmann distribution in the lower momentum limit. 
Power-law behavior of large $p_T$ distribution is also derived from the model.
} 
\maketitle
\section{Introduction}
\label{intro}
At RHIC colliding energy of nuclei becomes up to 200 AGeV, and thousands of particles 
are produced per event.  To describe such many particle system, 
a sort of collective approach will be useful. 
One-particle rapidity or pseudo-rapidity distributions observed at RHIC 
are well described by the Ornstein-Uhlenbeck process~\cite{biya04,wols99}.

In order to treat transverse momentum $p_T$ distributions, we can extend one dimensional stochastic equation
to a three-dimensional one using variables, longitudinal rapidity and two transverse momenta. 
The fundamental solution is Gaussian in these variables. 
Even if the formula is integrated over the azimuthal angle in the transverse momentum plane, 
the distribution is only slightly modified from the Gaussian in the low $p_T$ region. 
However, the observed $p_T$ distributions at RHIC have long tails compared with an exponential distribution, 
and cannot be described by the formula. 
 Therefore relativistic treatment for the transverse direction would be needed as long as the stochastic approach is adopted.
 
In reference~\cite{minh73}, an empirical formula for large $p_T$ distributions at polar angle $\theta=\pi/2$, 
\begin{eqnarray}
  E\frac{d^3\sigma}{d^3p}\left|_{\theta=\pi/2} \right. &=& A \exp\Big[-\frac{y_T^2}{2L_T}\Big],
    \nonumber \\
  y_T &=& \frac{1}{2}\ln \frac{E+|{\bf p}_T|}{E-|{\bf p}_T|},    \label{eq.int1}
\end{eqnarray}
was proposed from the analogy of Landau's hydrodynamical model. Polar angle $\theta$ ($0\le \theta \le \pi$) is measured from the beam direction of colliding nuclei or incident particles.  In equation (\ref{eq.int1}), $E$ denotes energy of an observed particle, $L_T$ is a parameter, and $y_T$ is called the "transverse rapidity".  Equation (\ref{eq.int1}) well describes large $p_T$ distributions for $p+p\rightarrow \pi^0 + X$ and $p+p\rightarrow \pi^\pm + X$.
Recently, large $p_T$ distributions for $\pi^0$ in $pp$ collisions observed at RHIC are also analyzed by (\ref{eq.int1})~\cite{stei04}.  However, it cannot be derived from the hydrodynamical model.  

The transverse rapidity is defined in the geodesic cylindrical coordinate system in the three dimensional rapidity space, where longitudinal rapidity $y$, transverse rapidity $\xi$ and azimuthal angle $\phi$ are used. 
The longitudinal rapidity $y$, and the transverse rapidity $\xi$ are defined respectively as,
 \begin{eqnarray*}
   y=\ln\frac{E+p_L}{m_T}, \quad \xi=\ln\frac{m_T+|{\bf p}_T|}{m},
 \end{eqnarray*}
where $E, p_L$, ${\bf p}_T$ and $m$ denote energy, longitudinal momentum, transverse momentum, and mass of the observed particle, respectively, and $m_T=\sqrt{{\bf p}_T^2 + m^2}$. It should be noted that $y_T$ coincides with $\xi$, only if $\theta=\pi/2$, namely $p_L=0$.
  
As for the relativistic approach to stochastic equation, we consider the diffusion equation in the three dimensional rapidity space or Lobachevsky space, which is non-Euclidean. In order to classify the longitudinal and transverse expansion, it would be appropriate to consider the diffusion equation in the geodesic cylindrical coordinate,
\begin{eqnarray}
  \frac{\partial f}{\partial t} &=& \frac{D}{\cosh^2\!\xi}\,\frac{\partial^2 f}{\partial y^2} 
         + \frac{D}{\sinh\xi \cosh\xi}\frac{\partial}{\partial \xi} 
          \left(\sinh\xi\cosh\xi\frac{\partial f}{\partial \xi}\right) \nonumber \\
         &&+ \frac{D}{\sinh^2\,\xi}\frac{\partial^2 f}{\partial \phi^2}, 
              \label{eq.int2}
\end{eqnarray}
where $D$ denotes a diffusion constant.  However, we cannot solve (\ref{eq.int2}) at present, as far as we are aware. Therefore, we should consider somewhat simpler case.

  We have proposed the relativistic diffusion model where an effect of radial flow is not included, and analyzed large $p_T$ distributions for charged particles~\cite{suzu04} and identified particles\cite{suzu04b} in $Au+Au$ collisions. 
  The distribution function of it is Gaussian-like in radial rapidity $\rho$, which coincides with $\xi$ at $\theta=\pi/2$, and resembles with (\ref{eq.int1}) at $\theta=\pi/2$.
  
  In section 2, the relativistic diffusion model, where the flow effect is taken into account, is briefly explained. 
  In section 3, large $p_T$ distributions in $p+p \rightarrow \pi^0 + X$ and in $Au+Au \rightarrow \pi^0 + X$ at 
  $\sqrt{s_{NN}}=200$ GeV observed at RHIC~\cite{adle03a,adle03b,adle03c} are analyzed.   
  In section 4, a flow effect or a transverse flow velocity is estimated from $p_T$ distributions in $Au + Au \rightarrow \bar{p} + X$, and $p_T$ distributions for $\pi^0, \pi^-$ and $K^-$ are analyzed by the use of the transverse flow velocity estimated from $p_T$ distributions of $\bar{p}$. 
  A possible interpretation of the dispersion of the Gaussian-like distribution is shown in section 5. The higher transverse momentum limit of our model is discussed in section 6.  
   Final section is devoted to summary and discussions.

\section{Diffusion equation in the three dimensional rapidity space}
\label{sec:1}
For simplicity, we consider the diffusion equation with radial symmetry in the three dimensional geodesic polar coordinate system,
\begin{eqnarray}
  \frac{\partial f}{\partial t}= \frac{D}{\sinh^2\!\rho}\, 
      \frac{\partial}{\partial \rho}\left( 
        \sinh^2\!\rho\, \frac{\partial f}{\partial \rho} 
      \right),           \label{eq.dif1}
\end{eqnarray}
with an initial condition,
\begin{eqnarray}
    f(\rho,t=0)= \frac{\delta(\rho-\rho_0)}
     {4\pi\sinh^2\!\rho}.     \label{eq.dif2}
\end{eqnarray}
In equation (\ref{eq.dif1}), $\rho$ denotes the radial rapidity, which is written with energy $E$, momentum ${\bf p}$ and mass $m$ of observed particle,  
\begin{eqnarray}
   \rho=\ln \frac{E+|{\bf p}|}{m}.    \label{eq.dif3}
\end{eqnarray}
Inversely, energy and momentum are written respectively as
\begin{eqnarray}
    E = m\cosh\rho,   \qquad
       |{\bf p}| = \sqrt{p_L^2+ {\bf p}_T^2} = m\sinh\rho.   \label{eq.dif4}
\end{eqnarray}

The solution~\cite{suzu04c} of  (\ref{eq.dif1}) with the initial condition (\ref{eq.dif2}) is given by
\begin{eqnarray}
  f(\rho,\rho_0,t) &=& \frac{1}{2\pi\sqrt{4\pi Dt}} {\rm e}^{-Dt}  \nonumber \\
    && \times\frac{\sinh{\frac{\rho_0\rho}{2\pi Dt}}}{\sinh\rho_0\sinh\rho}
     \exp \Big[ -\frac{\rho^2+\rho_0^2}{4Dt} \Big].   \label{eq.dif5a}
 \end{eqnarray}

A physical picture described by (\ref{eq.dif1}) and (\ref{eq.dif2}) is as follows;
After a collision of nuclei, particles are produced at radial rapidity $\rho_0$   expressed by (\ref{eq.dif2}). Then those particles propagate according to the diffusion equation (\ref{eq.dif1}).  In the course of the space time development, energy is supplied from the leading particle system (collided nuclei) to the produced  particle system. Number density of particles becomes lower and at some (critical) density, interactions among secondary particles cease, and particles become free.

From equation (\ref{eq.dif5a}), the following equation~\cite{karp59} is obtained;
\begin{eqnarray}
  f(\rho,t) &=& \lim_{\rho_0 \rightarrow 0}f(\rho,\rho_0,t)  \nonumber \\
            &=& \left( 4\pi Dt \right)^{-3/2} {\rm e}^{-Dt} \frac{\rho}{\sinh\rho}
     \exp \Big[ -\frac{\rho^2}{4Dt} \Big].   \label{eq.dif5b}
 \end{eqnarray}
Transverse momentum (rapidity) distributions at fixed polar angle $\theta$ are analyzed 
by (\ref{eq.dif5a}) or (\ref{eq.dif5b}), where transverse momentum is given
by $|{\bf p}_T|=m\sinh\rho\sin\theta$.

In order to clarify our approach to the previous non-relativistic approach~\cite{volo97,sasa01}, equation 
(\ref{eq.dif5a}) is compared with the solution for the radial symmetric diffusion equation 
in the Euclidean space in appendix A.

\section{Analysis of $p_T$ distributions observed at RHIC}
\label{sec:2}
At first, transverse momentum (rapidity) distributions of identified particles 
observed by the PHENIX collaboration~\cite{adle03a,adle03b,adle03c} are analyzed by the use of (\ref{eq.dif5b}), 
\begin{eqnarray}
  f(\rho,t)&=&  C \left(2\pi\sigma(t)^2\right)^{-3/2} 
      {\rm e}^{-\sigma(t)^2/2} \nonumber \\
      &&\times\frac{\rho}{\sinh\rho} 
     \exp\Big[-\frac{\rho^2}{2\sigma(t)^2} \Big], \nonumber \\
   \sigma(t)^2 &=& 2Dt,    \label{eq.dif6}
\end{eqnarray}
with parameters, $C$ and $\sigma(t)^2$
\footnote{Somewhat different analyses based on statistical models and the different stochastic model have been made in~\cite{biya05} }. 

As the data are taken at $\theta=\pi/2$ or $p_L=0$, the identity $\rho=\xi=y_T$ is satisfied.
In the analysis, we use $m_{\pi^0}=135$, $m_{\pi^-}=140$, $m_{K^-}=494$, and $m_{\bar{p}}=938$ MeV. 

\begin{figure}
  \begin{center}
    \includegraphics[scale=0.5,bb=30 220 480 650,clip]{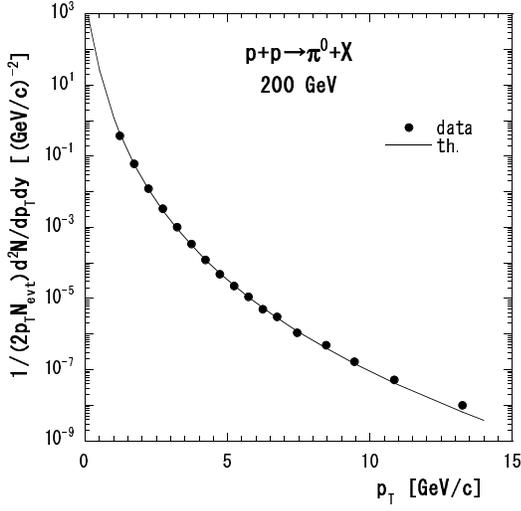}
  \end{center}
  \caption{\label{fig.pppi0}$p_T$ distribution for $p+p\rightarrow \pi^0+X$ at $y=0$ ~\cite{adle03a} }
 \end{figure}

The result on the $p_T$ distribution in $p+p \rightarrow \pi^0 +X$ is shown in Fig.~\ref{fig.pppi0}. Solid curve in Fig.~\ref{fig.pppi0} is drawn by the use of (\ref{eq.dif6}), parameters of which are estimated with the least square method, and are shown in Table~\ref{tab.table1}. Observed $p_T$ distributions for $\pi^0$ in $p+p$ collisions are well described by (\ref{eq.dif6})  from 1.2 GeV/c to 13.2 GeV/c.

\begin{figure}
 \begin{center}
    \includegraphics[scale=0.5,bb=30 160 530 650,clip]{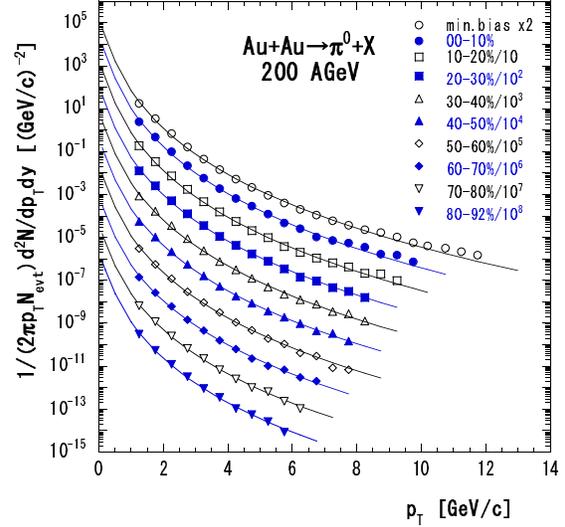}
 \end{center}
   \caption{\label{fig.aapi0}$p_T$ distribution for 
      $Au+Au\rightarrow \pi^0+X$ at $y=0$~\cite{adle03b} }
\end{figure}

  The results on $p_T$ distributions in $Au+Au\rightarrow\pi^0+X$ are shown in Fig.~\ref{fig.aapi0} and Table~\ref{tab.table2}.  As can be seen from Fig.~\ref{fig.aapi0} and Table~\ref{tab.table2},  observed $p_T$ distributions on $\pi^0$ in $Au+Au$ collisions are well described by (\ref{eq.dif6}) from centrality 0-10\% to centrality 80-92 \%.

\begin{table}
 \caption{ Parameters on $p_T$ distributions estimated by (\ref{eq.dif6})
  in $p+p \rightarrow \pi^0+X$  at $y=0$ at $\sqrt{s}=200$ GeV~\cite{adle03a}}
  \label{tab.table1}
 \begin{center}
 \begin{tabular}{cccc}  
  \hline\noalign{\smallskip}
     $C$                 & $\sigma(t)^2$    & $\chi^2$/n.d.f. \\ 
  \noalign{\smallskip}\hline\noalign{\smallskip}   
    13547.3 $\pm$ 941.7  & 0.603$\pm$ 0.004 & 12.3/15      \\ 
  \noalign{\smallskip}\hline
 \end{tabular}
 \end{center}
\end{table}
%
%
%
%
\begin{table}
 \caption{Parameters on $p_T$ distributions  estimated  by (\ref{eq.dif6}) 
   in $Au+Au\rightarrow \pi^0+X$ at $y=0$ at $\sqrt{ s_{NN}}=200$ GeV~\cite{adle03b}}
 \label{tab.table2} 
 \begin{center}
  \begin{tabular}{cccc} 
   \hline\noalign{\smallskip}
    centrality &  $C$  & $\sigma(t)^2$    & $\chi^2/$n.d.f  \\ 
   \noalign{\smallskip}\hline\noalign{\smallskip}
   min.bias &   45066.0 $\pm$  4542.7 & 0.586 $\pm$ 0.004 & 28.0/22 \\
   00-10\%  &  146819.0 $\pm$ 16095.3 & 0.574 $\pm$ 0.005 & 27.3/16 \\ 
   10-20\%  &  103772.0 $\pm$ 11348.3 & 0.580 $\pm$ 0.005 & 20.0/15 \\ 
   20-30\%  &   69568.0 $\pm$  7647.8 & 0.586 $\pm$ 0.005 & 14.9/13 \\ 
   30-40\%  &   43168.0 $\pm$  4870.7 & 0.592 $\pm$ 0.006 & 15.7/13 \\ 
   40-50\%  &   25292.0 $\pm$  2955.2 & 0.600 $\pm$ 0.006 &  9.4/12 \\ 
   50-60\%  &   13800.3 $\pm$  1629.2 & 0.600 $\pm$ 0.007 & 12.4/12 \\ 
   60-70\%  &    5477.2 $\pm$   699.8 & 0.617 $\pm$ 0.008 &  8.2/10 \\ 
   70-80\%  &   2498.80 $\pm$   336.74 & 0.616 $\pm$ 0.009 &  7.2/9  \\ 
   80-92\%  &   1275.89 $\pm$   216.03 & 0.610 $\pm$ 0.011 &  5.0/8  \\
  \noalign{\smallskip}\hline
 \end{tabular}
 \end{center}
\end{table}
%
%
\section{Estimation of radial flow}
\label{sec:3}
We have analyzed large $p_T$ distributions in $Au+Au\rightarrow \pi^0 +X$ in the previous section.  However, if the effect of flow in the transverse direction is taken into account, estimated values of parameters would be different from those shown in Table~\ref{tab.table2}.
  
\begin{figure}
 \begin{center}
    \includegraphics[scale=0.5,bb=30 160 530 650,clip]{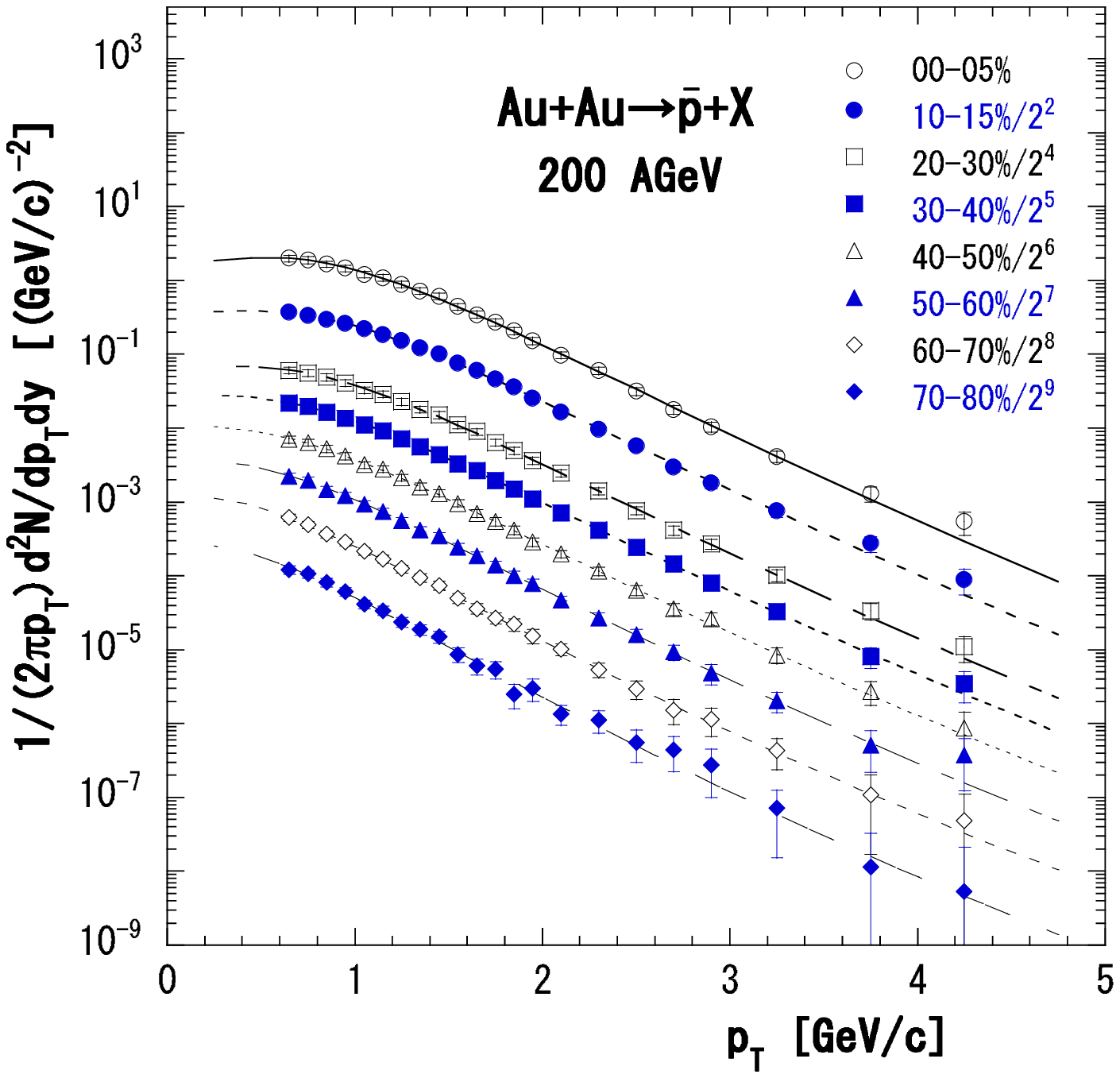}
 \end{center}
   \caption{\label{fig.pbarr0in} $p_T$ distribution for $Au+Au\rightarrow {\bar p}$ 
   at $y=0$~\cite{adle03b}, analyzed by (\ref{eq.flo1}), where $\rho_0$ is included. }
\end{figure}

\begin{figure}
 \begin{center}
    \includegraphics[scale=0.5,bb=60 160 520 650,clip]{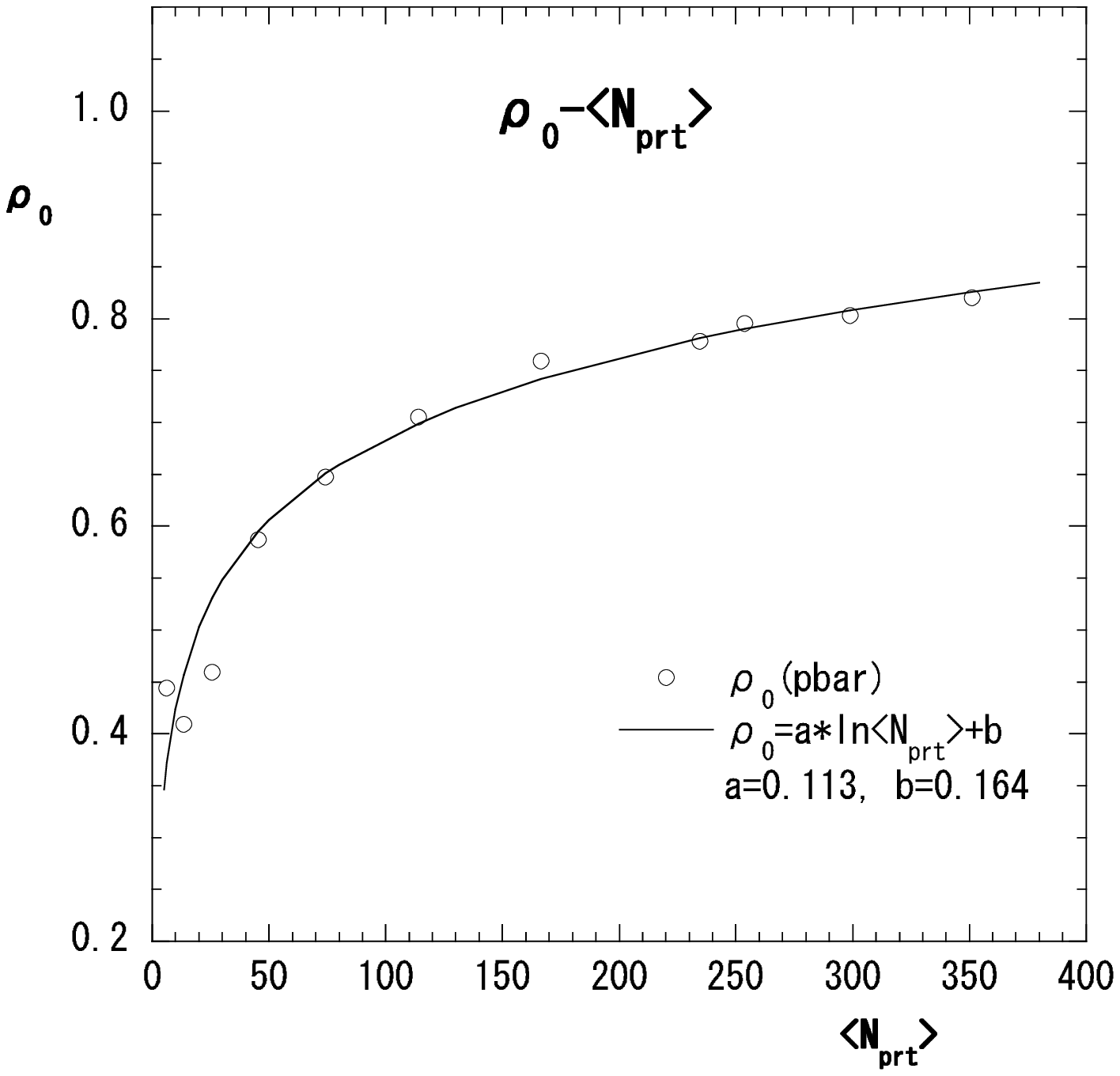} 
  \end{center}
  \caption{\label{fig.r0nprt} Dependence of $\rho_0$ to average number of participants in $Au+Au$ collisions
  estimated  by (\ref{eq.flo2}) }    
 \end{figure}

The effect of transverse flow in $Au+Au$ collisions at $\sqrt{s_{NN}}=200$ GeV is estimated from the observed $p_T$ distribution in $Au+Au \rightarrow \bar{p} + X$ from $p_T=0.65$ GeV/c to $p_T=4.25$ GeV/c 
by the use of the formula,
 \begin{eqnarray}
  f(\rho,\rho_0,t) &=& \frac{C}{2\pi\sqrt{2\pi\sigma(t)^2}} {\rm e}^{-\sigma(t)^2/2} \nonumber \\
   &&\times\frac{\sinh{\frac{\rho_0\rho}{\sigma^2(t)}}}{\sinh\rho_0\sinh\rho} 
     \exp \Big[ -\frac{\rho^2+\rho_0^2}{2\sigma(t)^2} \Big].      \label{eq.flo1}
 \end{eqnarray}
The results are shown in Fig.~\ref{fig.pbarr0in} and Table~\ref{tab.table3}. 
%
%
%
%
%
%
\begin{table*}
 \caption{Parameters on $p_T$ distributions estimated by (\ref{eq.flo1})  
    in $Au+Au\rightarrow \bar{p}+X$ at $y=0$ at $\sqrt{s_{NN}}=200$ GeV~\cite{adle03c}}
 \label{tab.table3} 
 \begin{center}
 \begin{tabular}{cccccc}  
  \hline\noalign{\smallskip} 
  centrality  & $C$ & $\sigma(t)^2$ & $\rho_0$ &$\chi^2/$n.d.f \\
  \noalign{\smallskip}\hline\noalign{\smallskip}
  00-05\%  & 17.713  $\pm$ 0.626  & 0.138$\pm$ 0.008 & 0.820$\pm$ 0.029 & 4.5/19\\ 
  05-10\%  & 14.974  $\pm$ 0.538  & 0.141$\pm$ 0.009 & 0.803$\pm$ 0.030 & 5.4/19\\ 
  10-15\%  & 12.508  $\pm$ 0.454  & 0.144$\pm$ 0.009 & 0.795$\pm$ 0.032 & 4.4/19\\ 
  15-20\%  & 10.492  $\pm$ 0.389  & 0.147$\pm$ 0.010 & 0.778$\pm$ 0.034 & 3.7/19\\ 
  20-30\%  &  7.762  $\pm$ 0.293  & 0.148$\pm$ 0.009 & 0.759$\pm$ 0.034 & 3.2/19\\ 
  30-40\%  &  5.175  $\pm$ 0.209  & 0.158$\pm$ 0.011 & 0.705$\pm$ 0.041 & 1.0/10\\ 
  40-50\%  &  3.168  $\pm$ 0.137  & 0.167$\pm$ 0.001 & 0.647$\pm$ 0.051 & 2.3/19\\ 
  50-60\%  &  1.8169 $\pm$ 0.0877 & 0.172$\pm$ 0.001 & 0.587$\pm$ 0.066 & 1.8/19\\ 
  60-70\%  &  0.8949 $\pm$ 0.0489 & 0.191$\pm$ 0.001 & 0.459$\pm$ 0.127 & 1.0/19\\ 
  70-80\%  &  0.3638 $\pm$ 0.0239 & 0.190$\pm$ 0.001 & 0.409$\pm$ 0.211 & 6.1/19\\ 
  80-92\%  &  0.1821 $\pm$ 0.0157 & 0.167$\pm$ 0.001 & 0.444$\pm$ 0.195 & 2.6/17\\
  \noalign{\smallskip}\hline
 \end{tabular}
 \end{center}
\end{table*}
%
%
From Table~\ref{tab.table3}, we can obtain the transverse flow velocity $v_0$ using the relation $v_0=\tanh \rho_0$. 
It varies from $v_0=\tanh 0.82 = 0.68$ at centrality 00-05\% to $v_0=0.42$ 
at centrality 80-92\% ($v_0=0.39$ at centrality 70-80\%).
As the relation of centrality to the average number of participant nucleons $\langle N_{prt} \rangle$ is given in Refs.~\cite{adle03a} and \cite{adle03b}, we can fit the estimated values of $\rho_0$ as a function of $\langle N_{prt} \rangle$ ;
 \begin{eqnarray}
  \rho_0=0.113\ln(x)+0.164, \quad x=\langle N_{prt} \rangle.  \label{eq.flo2}
 \end{eqnarray}
The result is shown in Fig.~\ref{fig.r0nprt}.

Then, we analyze observed $p_T$ distributions with $\rho_0$ given by (\ref{eq.flo2}), and $\langle N_{prt} \rangle$, which is shown in \cite{adle03b}. The results on $\pi^0$, $\pi^-$ and $K^-$  are shown respectively in Tables~\ref{tab.table2b}, \ref{tab.table3b} and \ref{tab.table4b}.

%
%
\begin{table*}
 \caption{Parameters on $p_T$ distributions estimated by (\ref{eq.flo1}) with (\ref{eq.flo2}) 
   in $Au+Au\rightarrow \pi^0+X$ at $y=0$ at $\sqrt{s_{NN}}=200$ GeV~\cite{adle03b}}
  \label{tab.table2b} 
 \begin{center}
 \begin{tabular}{ccccc}
  \hline\noalign{\smallskip} 
  centrality &  $C$  & $\sigma(t)^2$ &$\rho_0$  & $\chi^2/$n.d.f  \\ 
  \noalign{\smallskip}\hline\noalign{\smallskip}
   min.bias &  27977.0 $\pm$ 2764.2 & 0.500 $\pm$ 0.004 & 0.694 & 29.8/22 \\
   00-10\%  &  72773.0 $\pm$ 7625.2 & 0.469 $\pm$ 0.005 & 0.818 & 24.2/16 \\ 
   10-20\%  &  56883.0 $\pm$ 5936.0 & 0.479 $\pm$ 0.005 & 0.781 & 17.0/15 \\ 
   20-30\%  &  41647.0 $\pm$ 4386.1 & 0.490 $\pm$ 0.005 & 0.742 & 10.3/13 \\ 
   30-40\%  &  27513.0 $\pm$ 3012.5 & 0.503 $\pm$ 0.005 & 0.699 & 13.1/13 \\ 
   40-50\%  &  17515.0 $\pm$ 2002.8 & 0.518 $\pm$ 0.006 & 0.651 & 7.6/12 \\ 
   50-60\%  &  10559.0 $\pm$ 1224.0 & 0.526 $\pm$ 0.006 & 0.595 & 9.2/12 \\ 
   60-70\%  &  4548.40 $\pm$ 578.86 & 0.553 $\pm$ 0.008 & 0.531 & 7.2/10 \\ 
   70-80\%  &  2241.30 $\pm$ 313.24 & 0.564 $\pm$ 0.009 & 0.457 & 6.4/9  \\ 
   80-92\%  &  1178.50 $\pm$ 204.72 & 0.574 $\pm$ 0.011 & 0.372 & 4.9/8  \\ 
  \noalign{\smallskip}\hline
 \end{tabular}
 \end{center}
\end{table*}
%
%
\begin{table*}
 \caption{Parameters on $p_T$ distributions estimated by (\ref{eq.flo1}) wit (\ref{eq.flo2})  
  in $Au+Au\rightarrow \pi^-+X$ at $y=0$ at $\sqrt{s_{NN}}=200$ GeV~\cite{adle03c}}
 \label{tab.table3b} 
 \begin{center}
 \begin{tabular}{ccccc} 
  \hline\noalign{\smallskip} 
  centrality  & $C$ & $\sigma(t)^2$ & $\rho_0$ & $\chi^2/$n.d.f \\ 
  \noalign{\smallskip}\hline\noalign{\smallskip}
  00-05\%  & 32150.0 $\pm$ 846.3 & 0.533$\pm$ 0.003 & 0.826 & 357.4/26\\
  05-10\%  & 27161.0 $\pm$ 718.2 & 0.546$\pm$ 0.003 & 0.808 & 308.0/26\\ 
  10-15\%  & 22407.0 $\pm$ 594.1 & 0.559$\pm$ 0.004 & 0.790 & 271.1/26\\ 
  15-20\%  & 18795.0 $\pm$ 501.4 & 0.566$\pm$ 0.004 & 0.781 & 217.5/26\\ 
  20-30\%  & 14332.0 $\pm$ 383.4 & 0.582$\pm$ 0.004 & 0.742 & 195.3/26\\ 
  30-40\%  &  9337.0 $\pm$ 254.4 & 0.603$\pm$ 0.004 & 0.699 & 136.9/26\\ 
  40-50\%  &  5773.6 $\pm$ 161.0 & 0.617$\pm$ 0.004 & 0.651 & 108.6/26\\ 
  50-60\%  & 3365.10 $\pm$ 96.36 & 0.629$\pm$ 0.005 & 0.595 &  91.0/26\\ 
  60-70\%  & 1729.70 $\pm$ 51.22 & 0.640$\pm$ 0.005 & 0.531 &  80.8/26\\ 
  70-80\%  &  784.01 $\pm$ 24.37 & 0.655$\pm$ 0.006 & 0.457 &  54.4/26\\ 
  80-92\%  &  393.59 $\pm$ 12.88 & 0.664$\pm$ 0.007 & 0.372 &  39.9/26\\
  \noalign{\smallskip}\hline
 \end{tabular}
 \end{center}
\end{table*}
%
%
\begin{table*}
 \caption{Parameters on $p_T$ distributions estimated by (\ref{eq.flo1}) with (\ref{eq.flo2})  
    in $Au+Au\rightarrow K^-+X$ at $y=0$ at $\sqrt{s_{NN}}=200$ GeV~\cite{adle03c}}
 \label{tab.table4b}
 \begin{center}
  \begin{tabular}{ccccc}
   \hline\noalign{\smallskip} 
  centrality  & $C$ & $\sigma(t)^2$ & $\rho_0$ & $\chi^2/$n.d.f \\ 
   \noalign{\smallskip}\hline\noalign{\smallskip}
  00-05\%  & 254.63 $\pm$ 8.93 & 0.258$\pm$ 0.008 & 0.826 & 0.52/14\\ 
  05-10\%  & 212.58 $\pm$ 7.45 & 0.266$\pm$ 0.008 & 0.808 & 0.62/14\\ 
  10-15\%  & 175.59 $\pm$ 6.20 & 0.269$\pm$ 0.008 & 0.790 & 0.87/14\\ 
  15-20\%  & 145.12 $\pm$ 5.17 & 0.271$\pm$ 0.008 & 0.781 & 0.92/14\\ 
  20-30\%  & 108.69 $\pm$ 3.89 & 0.286$\pm$ 0.008 & 0.742 & 1.07/14\\ 
  30-40\%  & 70.370 $\pm$2.579 & 0.290$\pm$ 0.008 & 0.699 & 1.70/14\\ 
  40-50\%  & 42.070 $\pm$1.577 & 0.302$\pm$ 0.008 & 0.651 & 1.80/14\\ 
  50-60\%  & 23.404 $\pm$0.918 & 0.311$\pm$ 0.009 & 0.595 & 2.76/14\\ 
  60-70\%  & 11.308 $\pm$0.465 & 0.320$\pm$ 0.009 & 0.531 & 3.51/14\\ 
  70-80\%  &  4.782 $\pm$0.215 & 0.329$\pm$ 0.011 & 0.457 & 3.75/14\\ 
  80-92\%  &  2.274 $\pm$0.113 & 0.342$\pm$ 0.013 & 0.372 & 7.59/14\\ 
  \noalign{\smallskip}\hline
 \end{tabular}
 \end{center}
\end{table*}
%
%
 \begin{figure}
 \begin{center}
    \includegraphics[scale=0.45,bb=60 160 520 650,clip]{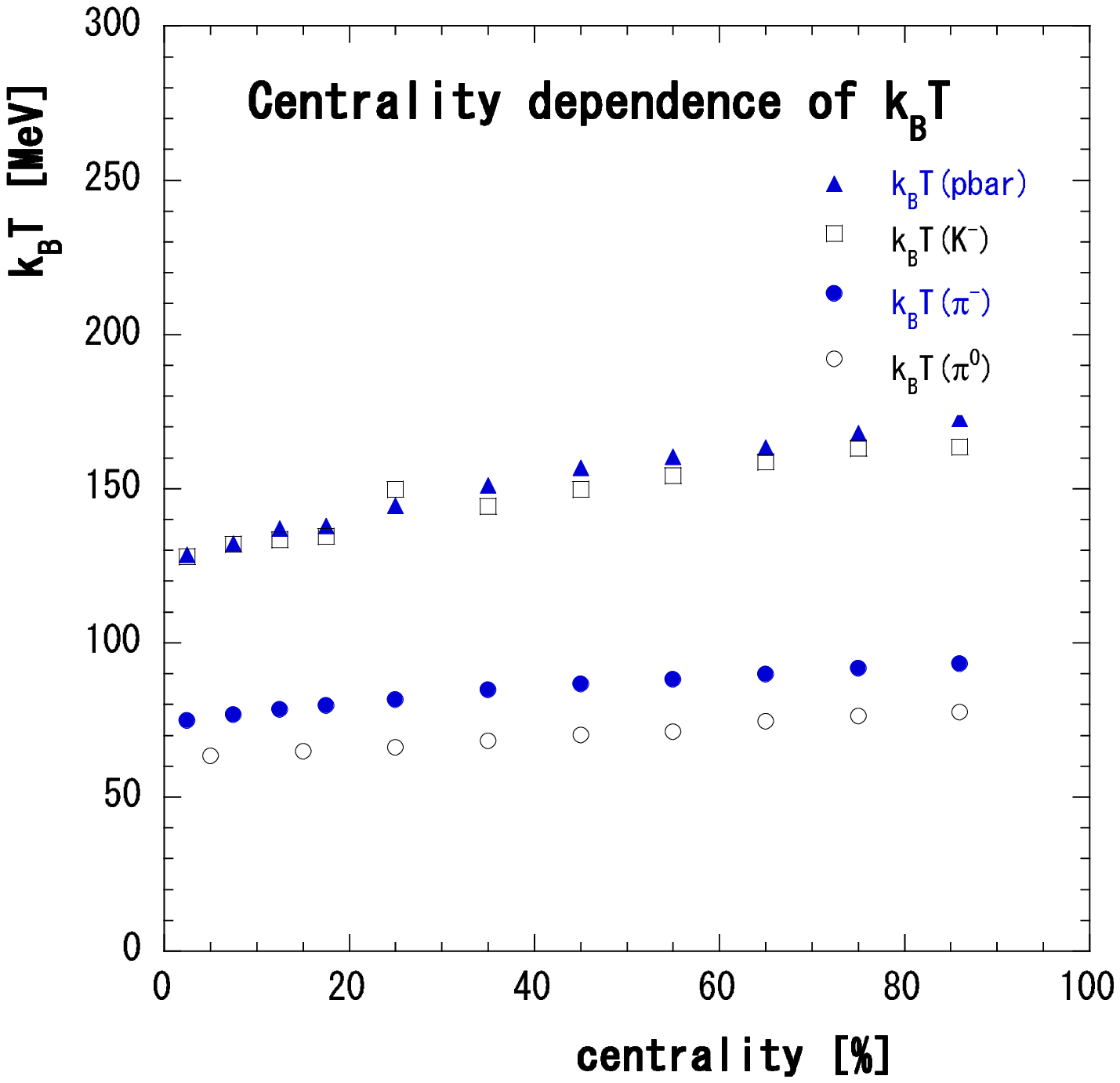} 
  \end{center}
  \caption{\label{fig.tempb} Centrality dependence of temperature $k_BT$ estimated
     by (\ref{eq.flo1}) 
    with (\ref{eq.flo2}) from $p_T$ distributions in $Au+Au$ collisions }
 \end{figure}

As can be seen from Tables~\ref{tab.table2b} and \ref{tab.table3b}, fitting of our calculation  by (\ref{eq.flo1}) to the data is better than that by (\ref{eq.dif6}) except for the minimum bias events in $Au+Au\rightarrow \pi^0 +X$. 
The fit to $\pi^-$ distributions is not so good as that to $\pi^0$ distributions. 
The result at lower centrality is no good, but it is much improved as the centrality increases. 

The results by (\ref{eq.flo1}) on $\pi^-$ and $K^-$ are also better than those by (\ref{eq.dif6}), 
which are not shown for brevity.

\section{Possible interpretation of parameter  $\sigma(t)^2$}
\label{section:4}
In order to get a possible interpretation of parameter $\sigma(t)^2$, an approximate expression of (\ref{eq.dif6}) in the small $|{\bf p}_T|$ region is taken.  
 When $\rho<<1$, $|{\bf p}|=m\sinh\rho\simeq m\rho$. Then, (\ref{eq.dif6}) reduces to
 \begin{eqnarray}
  f(\rho,t)\simeq \exp\Big[-\frac{\rho^2}{2\sigma(t)^2} \Big].   \label{eq.par1}
 \end{eqnarray}
If we assume that equation (\ref{eq.par1}) should coincide with the Maxwell-Boltzmann distribution, 
 \begin{eqnarray*}
  f({\bf v})=\left(\frac{m}{2\pi k_BT}\right)^{3/2}
        \exp\Big[-\frac{m{\bf v}^2}{2k_BT} \Big],   \label{eq.par2}
 \end{eqnarray*}
where $k_B$ is the Boltzmann constant, then we have an identity,
 \begin{eqnarray}
  k_BT=m\sigma(t)^2.   \label{eq.par3}
 \end{eqnarray}
Equation (\ref{eq.par3}) is obtained in the lower momentum limit, namely, with the condition that $\rho <<1$.

From (\ref{eq.par3}), we can estimate the temperature $k_BT$ of inclusive reactions for observed particle with mass $m$.  
The results are shown in Fig.~\ref{fig.tempb}. 
The estimated temperature $k_BT$ from the $p_T$ distributions for $\pi^0$ is from 63.3 MeV 
at centrality  5-10\% to 77.5 MeV at centrality from 80-92\%, and that for $\pi^-$ is from 74.8 MeV 
at centrality  0-5\% to 93.1 MeV at centrality from 80-92\%.
The temperature of $\pi^0$ or $\pi^-$, estimated from (\ref{eq.par3}), is somewhat lower than 
that from the exponential function of $p_T$.  
This would be affected by the assumption $\rho<<1$ used to obtain the identity (\ref{eq.par3}). 

The estimated temperature from the $p_T$ distributions for  $K^-$ is from 128.0 MeV at centrality  0-5\% to 163.5 MeV at centrality from 80-92\%.
That for ${\bar p}$ distributions is from 128.5 MeV at centrality  0-5\% to 172.6 MeV at centrality from 80-92\%. The results on $\bar{p}$ distributions shown in Fig.~\ref{fig.tempb} are recalculated by the use of (\ref{eq.par1}).

\section{Higher transverse momentum limit of the model}
\label{section:5}
In the 1970's, when large $p_T$ distributions are observed in accelerator experiments, many models, 
which have power-law behavior in $p_T$ are proposed~\cite{bjor74}. In \cite{hage83}, a model for $p_T$ distribution, inspired by QCD, is proposed as $(p_0/(p_T+p_0))^n$.
In the following, $|{\bf p}_T|$ is abbreviated to $p_T$.
It approaches an exponential distribution of $p_T$ for $p_T \rightarrow 0$, and a power function of $p_T$ for $p_T \rightarrow \infty$.

In this section, it is derived that equation (\ref{eq.dif5a}) at $\theta=\pi/2$ shows power-law behavior 
in $p_T$ in the higher $p_T$ limit, when the identity $\rho=\ln((m_T+p_T)/m)$ is used. 

The following two terms in (\ref{eq.dif5a}) can be rewritten in the power form of ${m_T+p_T}$ as;
\begin{eqnarray*}    
    \sinh{\frac{\rho_0\rho}{\sigma^2(t)}} &=& \frac{ 1-{\rm e}^{-2\rho_0\rho/\sigma^2(t)} }{2}
                     {\rm e}^{\rho_0\rho/\sigma^2(t)}  \nonumber  \\
    &&\simeq \frac{1}{2} \left(\frac{m_T + p_T}{m}\right)^{\rho_0/\sigma(t)^2},  \nonumber \\
    \sinh{\rho} &=& \frac{1-{ \rm e}^{-2\rho} }{2}{\rm e}^\rho
       \simeq \frac{1}{2}\left(\frac{m_T+ p_T}{m}\right).    
 \end{eqnarray*}
Radial rapidity contained in the Gaussian-like part in (\ref{eq.flo1}) is rewritten as,
\begin{eqnarray}
    \exp \Big[ -\frac{\rho^2}{2\sigma(t)^2} \Big] = \left(\frac{m_T + p_T}{m}\right)^{-\rho/{2\sigma(t)^2}}.  
              \label{eq.hig1}  
 \end{eqnarray}
As is well known that a logarithmic distribution of $x$ has weaker dependence compared 
with a power-law distribution of $x$ with any positive exponent in the limit of $x\rightarrow \infty$, 
we have, 
\begin{eqnarray*}
   \lim_{p_T\rightarrow\infty} \frac{\rho}{p_T^\epsilon}
    = \lim_{p_T\rightarrow\infty} \frac{1}{p_T^\epsilon} \ln\left(\frac{m_T+p_T}{m}\right) = 0
 \end{eqnarray*}
for any positive number $\epsilon$. Therefore, we can approximate $\rho$ in the exponent in (\ref{eq.hig1}) as constant, which is written by $2c_0$, within some finite transverse momentum range.  Then (\ref{eq.flo1}) is reduced to the form, 
\begin{eqnarray}
  f(\rho,\rho_0,t) &&\sim \left(\frac{m}{m_T+p_T}\right)^{(c_0-\rho_0)/{\sigma(t)^2}+1} \nonumber \\
      &&\sim \left(\frac{m}{2p_T}\right)^{(c_0-\rho_0)/{\sigma(t)^2}+1}. \label{eq.hig2}
 \end{eqnarray}
From equation (\ref{eq.hig2}), one can see that equation (\ref{eq.flo1}) shows power-law behavior in the higher transverse momentum limit, and that the power becomes smaller as $\rho_0$ increases, and it also becomes smaller as $\sigma(t)^2$,  which should increase with the colliding energy $\sqrt{s_{NN}}$, increases.

\section{Summary and discussions}
\label{section:6}
In order to analyze large $p_T$ distributions of charged particles observed at RHIC, the relativistic stochastic process in the three dimensional rapidity space, which is non-Euclidean, is introduced. The solution is Gaussian-like in radial rapidity, where the radial flow rapidity $\rho_0$ is included.  It is very similar to the formula proposed in \cite{minh73} at $\theta=\pi/2$.

Transverse momentum distributions for $\pi^0$ in $pp$ collisions at $y=0$ at $\sqrt{s}=200$ GeV observed by the PHENIX collaboration are analyzed. Observed $p_T$ distributions are well described by the formula (\ref{eq.dif6}) from 1.2 GeV/c to 13.2 GeV/c

In $Au+Au$ collisions, firstly, we analyzed the data for $\pi^0$ using (\ref{eq.dif6}), where flow effect is not included, i.e. $\rho_0=0$. The formula (\ref{eq.dif6}) well describes the observed $p_T$ distributions. 

Next, we have analyzed the observed $p_T$ distributions for $Au+Au\rightarrow {\bar p}+X$ using (\ref{eq.flo1}).  In the formula, non-zero radial rapidity $\rho_0$ at the initial stage, which corresponds to the radial flow velocity $v_0=\tanh\rho_0$, is included.

The temperature is estimated under the assumption that (\ref{eq.dif2}) approaches to the Maxwell-Boltzmann distribution when $\rho <<1$ or $p_T<<m$.

In our analysis, non-zero value of the radial flow rapidity (or radial flow velocity) has an influence 
on the estimation of temperature.  
One can see the effect of $\rho_0$ from (\ref{eq.hig2}).
In addition, all of the minimum chi-squared values by the use of (\ref{eq.flo1}) 
 with (\ref{eq.flo2}) become smaller than those with $\rho_0=0$ except for the minimum bias events in $Au+Au \rightarrow \pi^0 + X$.

We have derived that our formula (\ref{eq.flo1}) shows power-law behavior in $p_T$ in the higher transverse momentum limit. Therefore, it behaves like a Gaussian distribution in $p_T$ when $p_T<<m$, and like a power-law distribution 
in $p_T$ when $p_T>>1$.

\section*{Acknowledgement}
 Authors would like to thank RCNP at Osaka University, Faculty of Science, Shinshu University, 
 and Matsumoto University for financial support.

%
\appendix
\section{Relation to non-relativistic approach}

The radial symmetric diffusion equation in the $n$ ($n=2,3$) dimensional Euclidean space 
is written as,
\begin{eqnarray}
  \frac{\partial f}{\partial t}=  D\frac{\partial^2 f}{\partial r^2}  
   + \frac{D(n-1)}{r}\frac{\partial f}{\partial r}.
         \label{eq.nr1}
\end{eqnarray}

The solution of (\ref{eq.nr1}) with the initial condition
\begin{eqnarray}
  f(r,t=0)=\frac{\delta(r-r_0)}{2\pi(2r_0)^{n-1} }       \label{eq.nr2}
\end{eqnarray}
is given by~\cite{fell51},
\begin{eqnarray}
   f(r,r_0,t) &=& \frac{1}{2Dt}   
           \exp \Big[ -\frac{r^2+r_0^2}{4Dt} \Big]   \nonumber \\
         &&\times \left( {r_0}{r} \right)^{-(n-2)/2}
               I_{(n-2)/2} \left( \frac{r_0r}{2Dt}\right ). \label{eq.nr3}
\end{eqnarray}
In the case of two dimensional expansion, $n=2$, (\ref{eq.nr3}) coincides with the formula in \cite{volo97}, 
if $2Dt={m}/T$, $r=|{\bf p_T}|/m$, and $r_0=|{\bf p_{T0}}|/m$. 

Using the identity 
 \begin{eqnarray*}
    \sinh z = \sqrt{\pi z/2}\, I_{1/2}(z),
 \end{eqnarray*}
we can rewrite (\ref{eq.dif5a}) as
\begin{eqnarray}
  f(\rho,\rho_0,t) &=& \frac{1}{8\pi Dt}{\rm e}^{-Dt}
  \frac{ (\rho_0\rho)^{1/2} }{ \sinh\rho_0\sinh\rho }   \nonumber \\
  &&\times   \exp \Big[ -\frac{\rho^2+\rho_0^2}{4Dt} \Big] I_{1/2}\left(\frac{\rho_0\rho}{2Dt}\right).   
      \label{eq.nr5}
 \end{eqnarray}
In the case of the three dimensional expansion in the Euclidean space, comparing (\ref{eq.nr5}) 
with (\ref{eq.nr3}) at $n=3$, one can recognize that (\ref{eq.nr5}) is a relativistic extension 
of  (\ref{eq.nr3}), and that (\ref{eq.nr5}) reduces to (\ref{eq.nr3}) in the low momentum region, $|{\bf p}|<<1$, where $|{\bf p}|=m \sinh \rho \simeq m\rho$ and $\sinh \rho_0 \simeq \rho_0$.
  
%

%
\end{document}